\documentclass[12pt, 
]{article}

\usepackage{graphicx}
\usepackage{amsfonts}
\usepackage{amsmath}
\usepackage[usenames]{color}
\usepackage{amssymb}
\usepackage[mathscr]{eucal} 
\usepackage{longtable}

\usepackage{color}
\input{colordvi.tex}

\def\kesu#1{}

\def\N{\mathbb{N}}

\def\P{\mathbb{P}}

\newcommand{\sfrac}[2]{\left(\frac{#1}{#2}\right)}

\begin{document}
\baselineskip 0.6cm
\newcommand{\gsim}{ \mathop{}_{\textstyle \sim}^{\textstyle >} }
\newcommand{\lsim}{ \mathop{}_{\textstyle \sim}^{\textstyle <} }
\newcommand{\vev}[1]{ \left\langle {#1} \right\rangle }
\newcommand{\bra}[1]{ \langle {#1} | }
\newcommand{\ket}[1]{ | {#1} \rangle }
\newcommand{\Dsl}{\mbox{\ooalign{\hfil/\hfil\crcr$D$}}}
\newcommand{\nequiv}{\mbox{\ooalign{\hfil/\hfil\crcr$\equiv$}}}
\newcommand{\nsupset}{\mbox{\ooalign{\hfil/\hfil\crcr$\supset$}}}
\newcommand{\nni}{\mbox{\ooalign{\hfil/\hfil\crcr$\ni$}}}
\newcommand{\EV}{ {\rm eV} }
\newcommand{\KEV}{ {\rm keV} }
\newcommand{\MEV}{ {\rm MeV} }
\newcommand{\GEV}{ {\rm GeV} }
\newcommand{\TEV}{ {\rm TeV} }

\def\diag{\mathop{\rm diag}\nolimits}
\def\tr{\mathop{\rm tr}}

\def\Spin{\mathop{\rm Spin}}
\def\SO{\mathop{\rm SO}}
\def\O{\mathop{\rm O}}
\def\SU{\mathop{\rm SU}}
\def\U{\mathop{\rm U}}
\def\Sp{\mathop{\rm Sp}}
\def\SL{\mathop{\rm SL}}

\def\change#1#2{{\color{blue}#1}{\color{red} [#2]}\color{black}\hbox{}}


\begin{titlepage}
  
\begin{flushright}
  UT-11-14 \\
  IPMU11-0080 \\
  NSF-KITP-11-075 \\
\end{flushright}
   
\vskip 1cm
\begin{center}
  
 {\large \bf Investigating Generalized Parton Distribution in Gravity Dual}

\vskip 1.2cm
   
Ryoichi Nishio$^{1,2}$ and Taizan Watari$^2$
    
\vskip 0.4cm
 
{\it
  $^1$Department of Physics, University of Tokyo, Tokyo 113-0033, Japan  
   \\[2mm]
   
  $^2$Institute for the Physics and Mathematics of the Universe, University of Tokyo,\\ Chiba 277-8583, Japan
   }
\vskip 1.5cm
   
\abstract{
Generalized parton distribution (GPD) contains rich information of partons in a hadron,
including transverse profile,
and is also non-perturbative information necessary in describing a variety of hard processes, such as meson leptoproduction 
and double deeply virtual Compton scattering (DDVCS).
In order to unveil non-perturbative aspects of GPD,
we study DDVCS at small $x$ in gravitational dual description.
Using the complex spin $j$-plane representation of DDVCS amplitude,
we show that GPD is well-defined and can be extracted from the amplitude even in the strong coupling regime.
It also turns out that the saddle point value in the $j$-plane representation plays an important role;
there are two phases in
the imaginary part of amplitude of DDVCS and GPD,
depending on relative position of the saddle point and
the leading pole in the $j$-plane,
and crossover between them is induced by the change of the kinematical variables.
The saddle point value also directly controls kinematical variable dependence of many observables
in one of the two phases,
and indeed the dependence is qualitatively in nice agreement with HERA measurements.
Such observation that the gravity dual shares basic properties of the real world QCD
suggests that 
information from BFKL theory might be used to reduce error in the gravity dual predictions 
of the form factor and of GPD.
This article also serves as a brief summery of a preprint arXiv:1105.2999.
} 
   
\end{center}
\end{titlepage}



\section{Introduction}

AdS/CFT correspondence and its extension to non-conformal theories 
have been exploited for study of non-perturbative aspects of strongly 
coupled gauge theories. Hadron spectra, coupling constants among them
and chiral symmetry breaking have been studied intensively in the
literature by using gravitational dual descriptions with smooth infra-red 
non-conformal geometries. The gravitational dual approach 
can be used, however,
to study not just {\it static} properties of strongly coupled 
gauge theories, but also {\it scattering} of hadrons. Indeed, string theory 
or dual resonance model was originally constructed to describe
scattering of hadrons. Qualitative aspects of hadron scattering can be 
obtained in gravitational dual descriptions, if the background geometry 
(target space) of string theory is chosen properly \cite{Polchinski2002,
PolchinskiJHEP0305:0122003, BrowerJHEP0712:0052007}.

In this article, we will study 2-body to 2-body scattering of a hadron and
a virtual photon at high energy in gravitational dual descriptions.
 This process is called double deeply virtual Compton scattering (DDVCS). When
the final sate photon is on-shell, it is called deeply virtual Compton 
scattering (DVCS), and is accessible in experiments \cite{DVCS-expr-A}. 
Because of QCD factorization theorem \cite{factorization}, 
the DVCS or DDVCS amplitude 
is obtained as a convolution of generalized parton distribution (GPD)
\cite{GPD-original} and a hard kernel, the latter of which can be calculated 
in perturbative QCD. 
GPD itself (at a certain factorization scale), however, is a 
non-perturbative object in nature, and cannot be calculated in 
perturbative QCD. Even in determining it by using experimental data, 
its profile needs to be parametrized\footnote{See \cite{GPD-review} for 
review articles, which also have extensive list of literatures.} 
based on proper understanding on non-perturbative dynamics behind 
confinement. We thus use gravitational dual descriptions to extract 
theoretical understanding on the GPD profile. 

It is not that we just use a well-developed technique to calculate 
a specific scattering amplitude (or GPD) in this article, however. 
This article clarifies structure of Pomeron ``exchange'' amplitudes, 
how to organize them, as well as their field-theory interpretation.
We find that a saddle point value of the scattering amplitude in 
complex spin $j$-plane representation is a key concept in organizing 
Pomeron amplitudes and in understanding kinematical variable dependence 
of the scattering amplitude.
Based on this understanding, sharp cross-over behavior 
is expected in the photon-hadron 2-to-2 scattering amplitude 
in small $x$ limit.

This article is meant to be a brief summary of reference \cite{NW}.  
To keep this letter short enough, we extracted material mainly from 
\S 5 of \cite{NW}, and only minimum from other sections, imagining 
people in perturbative QCD community as primary readers of this letter. 
More theoretical aspects of the scattering amplitude in gravity dual,
as well as more detailed account of the materials in this letter,
are found in \cite{NW}.

\section{Amplitude in Gravity Dual}

In order to calculate hadron--virtual photon scattering amplitude 
in gravitational dual, one needs to adopt a certain holographic model. 
Since the real world QCD turns from weak coupling at high energy into 
strong coupling at infrared, it is desirable to have a holographic model 
that is faithful to string theory where AdS curvature becomes larger
than string scale toward UV boundary. Such a model becomes even more 
realistic, if spontaneous chiral symmetry breaking is implemented in
it. Our primary goal in this article, however, is not in pursuing 
precision in numerical calculation (as lattice QCD does) by 
setting up a perfectly realistic gravitational dual description. 
An appropriate set-up that suits the best for one's purpose should 
depend on the purpose. 

We will focus on qualitative aspects of hadron--virtual photon scattering 
amplitude at small $x$ (at high center-of-mass energy). Since small $x$ 
physics is dominated by gluon, not by quarks and anti-quarks, we do not 
find it a crucial element to implement flavor in the gravitational set
up for the purpose of this article. For explicit calculation, we adopt the
hard wall model \cite{PolchinskiJHEP0305:0122003}, which is type IIB string
theory on $W \times {\rm AdS}_5$ for some 5-dimensional manifold $W$ 
with ${\rm AdS}_5$ cut off at finite radius at infrared. Such a crude 
treatment of infrared geometry is sufficient for our qualitative 
study \cite{PolchinskiJHEP0305:0122003}, and the choice of $W$ becomes 
irrelevant (at least directly) for sufficiently small $x$ \cite{NW}. 
Since it is almost straightforward to see how the $AdS_5$ curvature and 
running of dilaton expectation value affects various observables 
in explicit calculations based on the hard wall model, one can also 
learn what happens in gravitational dual models that are asymptotically 
conformal or asymptotically free without carrying out calculations 
separately on these models.

As an analogy of the electromagnetic global $\U(1)$ symmetry of QCD, 
we take a global symmetry of $W$ in the gravitational dual. Since we are 
interested in the Compton tensor\footnote{In our convention, $\eta_{\mu\nu}=\text{diag}(-,+,+,+)$.
Let us remark that 
the Compton tensor $T^{\mu\nu}$ in this letter is defined differently from one in \cite{NW};
the Lorentz indices $\mu, \nu$ are interchanged.
}
of QCD, 
\begin{equation}
 i (2\pi)^4 \delta^4(p_2 + q_2 - p_1 - q_1) T^{\mu\nu} = - \int \int 
  d^4x d^4y e^{- i q_2 \cdot x} e^{+ i q_1 \cdot y} 
  \bra{h(p_2)}T\{ J^\mu(x)J^\nu(y)\}\ket{h(p_1)}, 
\end{equation}
we use the bulk-to-boundary propagator of an $AdS_5$ vector field 
associated with a Killing vector of $W$ in calculating the matrix
element involving the global symmetry current. As for the target  
hadron in the gravity dual, we use a Kaluza--Klein state of a dilaton, 
whose wavefunction is given by a Bessel function in the hard wall model.  
Thus, the leading order contribution in $1/N_c$ expansion is given by 
a closed string sphere amplitude with four NS--NS string vertex operator 
insertions \cite{PolchinskiJHEP0305:0122003}.\footnote{
The target hadron which is dual to a Kaluza--Klein state of a dilaton is a glueball.
The case of a meson target can also be studied in the same way
if we use open strings.
For the case of a baryon target, we should use $D$-brane in the gravity dual.
We will see 
that the saddle point value and singularities in the complex $j$-plane
representation are important in describing the amplitude.
Because they do not depend on the target hadron wavefunctions,
they are expected to be unchanged even if the species of target hadron is replaced.
}

As we consider cases where the initial state ``photon'' or both the initial 
and final state ``photons'' are highly virtual, that is, 
$q_1^2 \gg \Lambda^2$ or $q_1^2, q_2^2 \gg \Lambda^2$, 
the ``photon''--hadron scattering amplitude $T^{\mu\nu}$ can be 
decomposed into various contributions through operator product
expansion of $J^\mu(x)$ and $J^\nu(y)$ in QCD language. Such a
decomposition still holds true in strongly coupled gauge theories 
(and hence in gravitational dual), except that the anomalous dimensions 
of operators in the expansion may be quite different from what one
expects in the weak coupling regime. 
Reference \cite{PolchinskiJHEP0305:0122003} noted that the operators 
that are twist-2 in the weakly coupled regime still appear in the 
operator product expansion even in the strongly coupled regime, and their 
contributions to the Compton tensor $T^{\mu\nu}$ dominate at
sufficiently small $x$; this is because the ``twist-2'' contribution 
corresponds to exchange of leading Regge trajectory containing graviton 
in gravity dual language \cite{Gubser:2002tv, BrowerJHEP0712:0052007}.
We will thus focus on small $x$ hadron--virtual photon scattering 
in gravity dual to study non-perturbative behavior of the ``twist-2''
contribution. 

Before writing down the Pomeron contribution to the scattering 
amplitude explicitly, let us note that the Compton tensor 
is described by five structure functions $V_{1,2,\cdots,5}$ as in 
\cite{Marquet2010},
\begin{align}
 T^{\mu\nu} = &V_1 P[q_2]^{\mu\rho} P[q_1]^{\nu}_{\rho}
           +V_2 (p\cdot P[q_2])^\mu (p\cdot P[q_1])^\nu
           +V_3 (q_1\cdot P[q_2])^\mu (q_2\cdot P[q_1])^\nu
\notag \\ \label{eq:structure functions of Compton tensor}
           &+V_4 (q_1\cdot P[q_2])^\mu (p\cdot P[q_1])^\nu
           +V_5 (p\cdot P[q_2])^\mu (q_2\cdot P[q_1])^\nu
           -A \epsilon^{\mu\nu\rho\sigma}q_{1\rho}q_{2\sigma},
\end{align}
for a scalar target hadron, because of gauge invariance. 
In parity-preserving theory, $A=0$. In the limit 
of purely forward scattering, the two structure functions of 
deep inelastic scattering are restored from 
${\rm Im} \; V_1 (x,\eta, t, q^2) \rightarrow F_1(x,q^2)$ and 
$(q^2/(2x)) \times {\rm Im} \; V_2 (x,\eta, t, q^2) \rightarrow F_2(x,q^2)$.
Here, we introduced a convenient notation 
\begin{align}
  P[q]_{\mu\nu}=\left[\eta_{\mu\nu}-\frac{q_\mu q_\nu}{q^2}\right].
\end{align}
In this article, we will use the following notations, 
\begin{equation}
 q^\mu = \frac{(q_1+q_2)^\mu}{2}, \quad 
 p^\mu = \frac{(p_1 + q_2)^\mu}{2}, \quad 
 x = - \frac{q^2}{2 p \cdot q}, \qquad 
 \eta = - \frac{q \cdot (q_1 - q_2)}{2 p \cdot q}, 
\end{equation}
and $t = - (q_1 - q_2)^2$ and $s = W^2 = - (q + p)^2$.

In the generalized Bjorken limit, $\Lambda^2, |t| \ll q_1^2$, and for  
$x$ much smaller than unity, the Pomeron contribution to the five structure
functions are given by $I_0$ and $I_1$ \cite{NW} as in 
\begin{align}
 V_1&\simeq \frac{1}{2} I_1, &
 V_2&\simeq \frac{2x^2}{q^2}(I_0+I_1), & 
 V_3&\simeq \frac{x^2}{2q^2}(I_0+I_1), \notag \\
 V_4&\simeq \frac{x}{q^2}I_1,&V_5&\simeq \frac{x}{q^2}I_1;& & 
\label{eq:DVCS-polarization}
\end{align}
$I_0$ and $I_1$ are given for vanishing skewedness $\eta$ in the form of  
\begin{equation}
 I_i (x, \eta, t, q^2) \simeq \frac{c'_s}{2 \kappa_5^2}\frac{\pi}{2 R^3} 
  \int dz \sqrt{-g(z)} \int dz' \sqrt{-g(z')} 
   P^{(i)}_{\gamma^\ast\gamma^\ast}(z) \; {\cal K}(s,t,z,z') \; 
   P_{hh}(z').
\label{eq:Ii-def}
\end{equation}
For vanishing skewedness,
 the Pomeron kernel ${\cal K}$
 is \cite{BrowerJHEP0712:0052007}\footnote{
More careful discussion on the choice of integration contour is given in \cite{Hatta:2007he, NW}.
A pedagogical explanation of the origin of $1/\Gamma^2(j/2)$ factor is also given in \cite{NW}.
}
\begin{align}
 {\cal K}(s,t;z,z') \simeq
& -4R\sqrt{\lambda}
 \int_{-\infty}^\infty d\nu \frac{1}{2\pi i}  \int_{C_1(\nu)}dj \; 
  \frac{1+e^{-i\pi j}}{\sin\pi j} \frac{1}{\Gamma^2(j/2)} \notag \\
\label{eq:pomeron kernel}
& \sfrac{\alpha'\tilde s}{4}^{j}
\frac{1}{j-j_{\nu}} \; 
e^{-j A(z)}\Psi^{(j)}_{i\nu}(t,z) \; e^{-j A(z')}
 \Psi^{(j)}_{i\nu}(t,z');
\end{align}
the integration contour in the complex $j$-plane encircles the pole 
$j = j_\nu$, and once the residue of this pole is picked up, a relation
\begin{equation}
 j = j_\nu \equiv 2 - \frac{4+\nu^2}{2\sqrt{\lambda}}
\label{eq:j-jnu}
\end{equation}
sets the (analytically continued) relation between spin $j$ and 
anomalous dimension $\gamma = i\nu - j$ of ``twist-2'' operators 
in the large 't Hooft coupling $\lambda \gg 1$ 
regime \cite{BrowerJHEP0712:0052007}.
$e^{2A(z)} = (R/z)^2$ is the warp factor in the ${\rm AdS}_5$ 
part of the metric in the hard wall model, 
\begin{equation}
 ds^2|_{{\rm AdS}_5} = e^{2A(z)} (\eta_{\mu\nu} dx^\mu dx^\nu + (dz)^2),  
\end{equation}
and $\sqrt{-g}$ in (\ref{eq:Ii-def}) is that of this metric of
5-dimensional spacetime. $R$ is the AdS radius, and the infrared cut off 
of the hard wall model at $z = 1/\Lambda$ sets the confinement scale $\Lambda$.
$\tilde{s} = e^{-A(z)}e^{-A(z')} s$, and $\alpha'$ is the slope
parameter of the Type IIB string theory. 
$\Psi^{(j)}_{i\nu}(t,z)$ in (\ref{eq:pomeron kernel}) is the Pomeron 
wavefunction in the spin $j$ channel, which is given by 
\begin{align}\label{psi}
 \Psi^{(j)}_{i\nu}(t,z)=ie^{A(j-2)}\sqrt{\frac{\nu}{2R\sinh\pi\nu}}\left[
\sqrt{\frac{ I_{-i\nu} (\sqrt{-t}/\Lambda)      }{I_{i\nu} (\sqrt{-t}/\Lambda)}}I_{i\nu}(\sqrt{-t}z)-
\sqrt{\frac{ I_{ i\nu} (\sqrt{-t}/\Lambda)      }{I_{-i\nu} (\sqrt{-t}/\Lambda)}}I_{-i\nu}(\sqrt{-t}z)
\right]
\end{align}
in the hard wall model.\footnote{Dirichlet boundary condition was
imposed at the infrared boundary $z = 1/\Lambda$, just to make expressions
simpler.} 

The impact factor $P_{hh}(z')$ of the target hadron side is given by 
the normalizable mode wavefunction of the target hadron, as in 
$P_{hh}(z') = c_{\phi} (\Phi(z'))^2$. On the virtual ``photon'' side, 
the bulk-to-boundary propagator (non-normalizable wavefunction) 
of the graviton associated with the Killing vector of $W$ is used; 
in the hard wall model, they are 
\begin{eqnarray}
 P^{(1)}_{\gamma^\ast\gamma^\ast}(z) & = & c_J^2 R^2 e^{-2 A(z)}
   [(q_1z)(K_1(q_1z)][(q_2 z)K_1(q_2 z)], \label{eq:P1-def}\\
P^{(0)}_{\gamma^\ast\gamma^\ast}(z) & = & \frac{c_J^2 R^2 e^{-2 A}}{q^2}
   [(q^2_1z)(K_0(q_1z)][(q^2_2 z)K_0(q_2 z)]  \label{eq:P0-def}
\end{eqnarray}
for $I_1$ and $I_0$, respectively. $\kappa_5^2$ is a constant of a
theory of mass dimension $-3$ and is proportional to $N_c^2$. 
$c_s'$, $c_\phi$ and $c_J$ are dimensionless constants of order unity. 
See \cite{NW} for their definitions. 

\section{Structure and Behavior of the Amplitude}

\subsection{Complex $j$-plane amplitude, Pomeron vertex and form factor}

\begin{figure}[tbp]
  \begin{center}
\begin{tabular}{cccc}
  \includegraphics[scale=0.23]{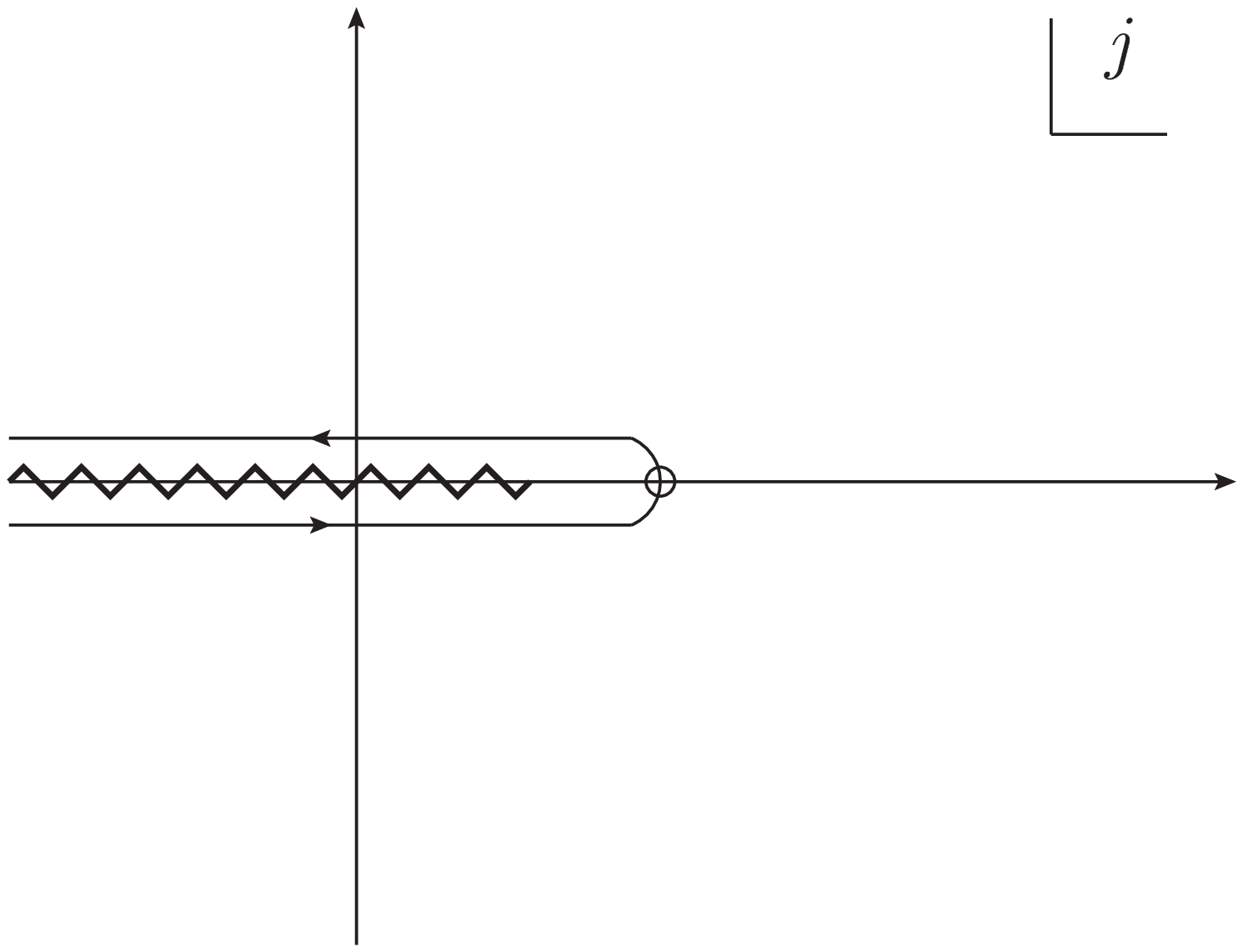} & 
  \includegraphics[scale=0.23]{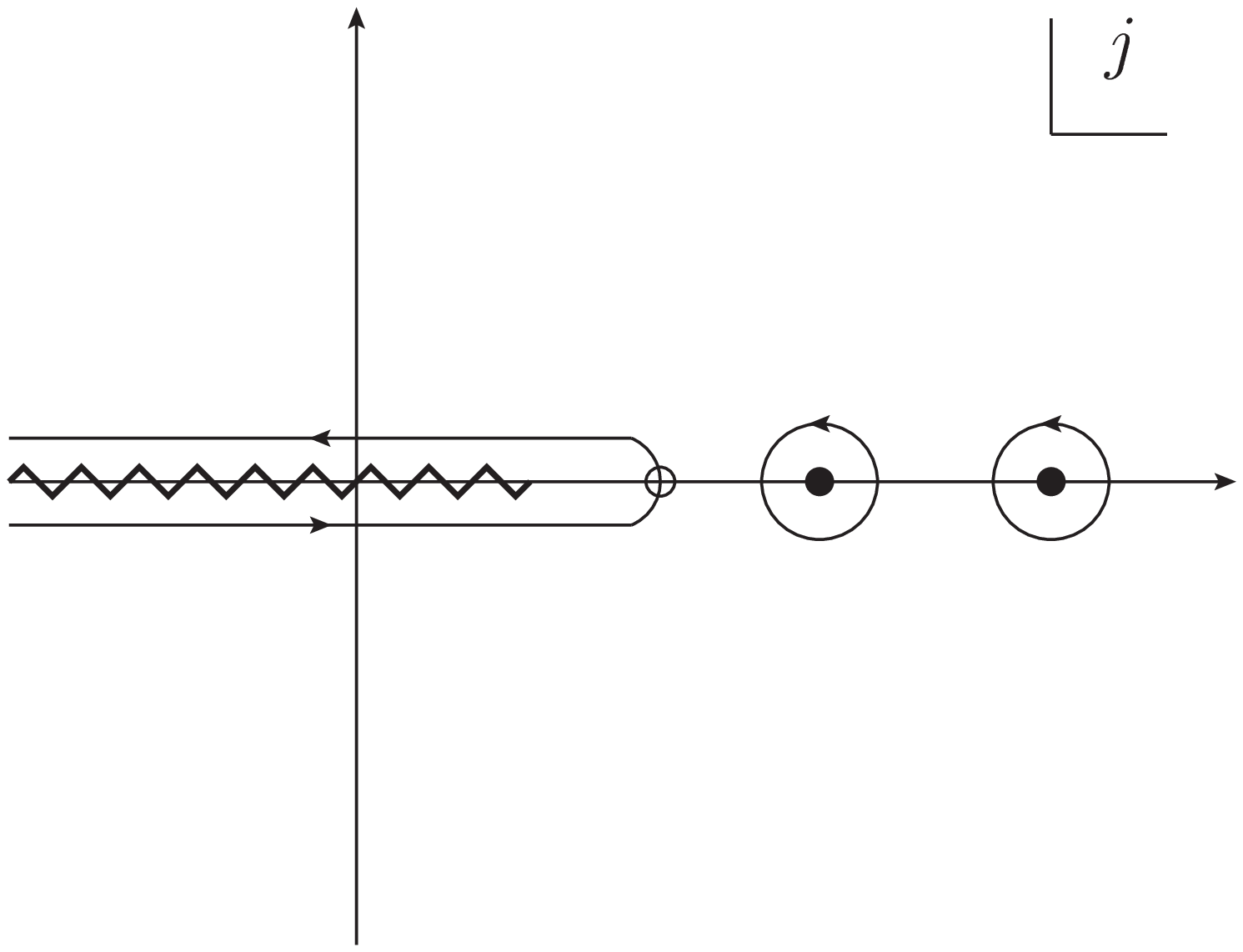} &
  \includegraphics[scale=0.23]{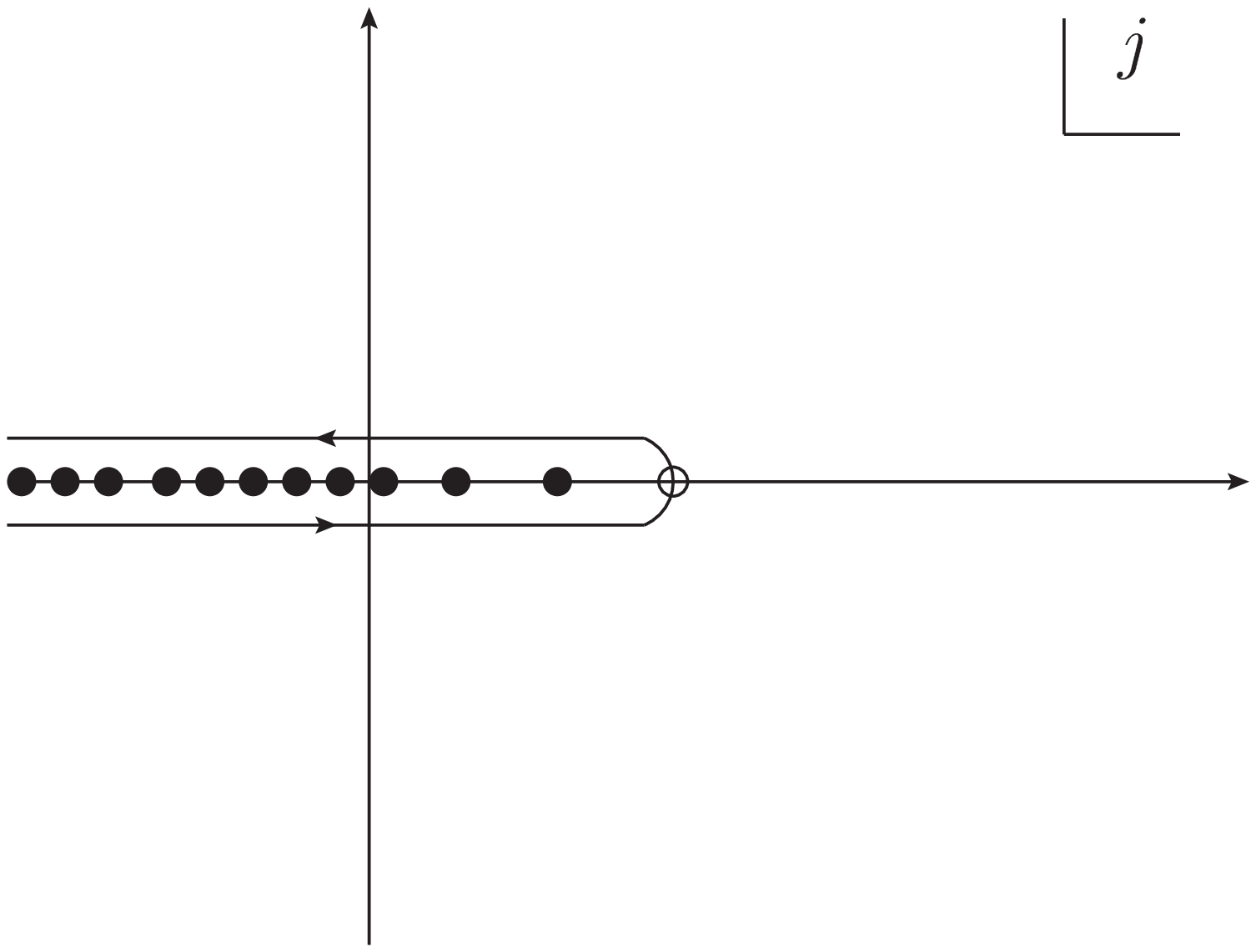} & 
  \includegraphics[scale=0.23]{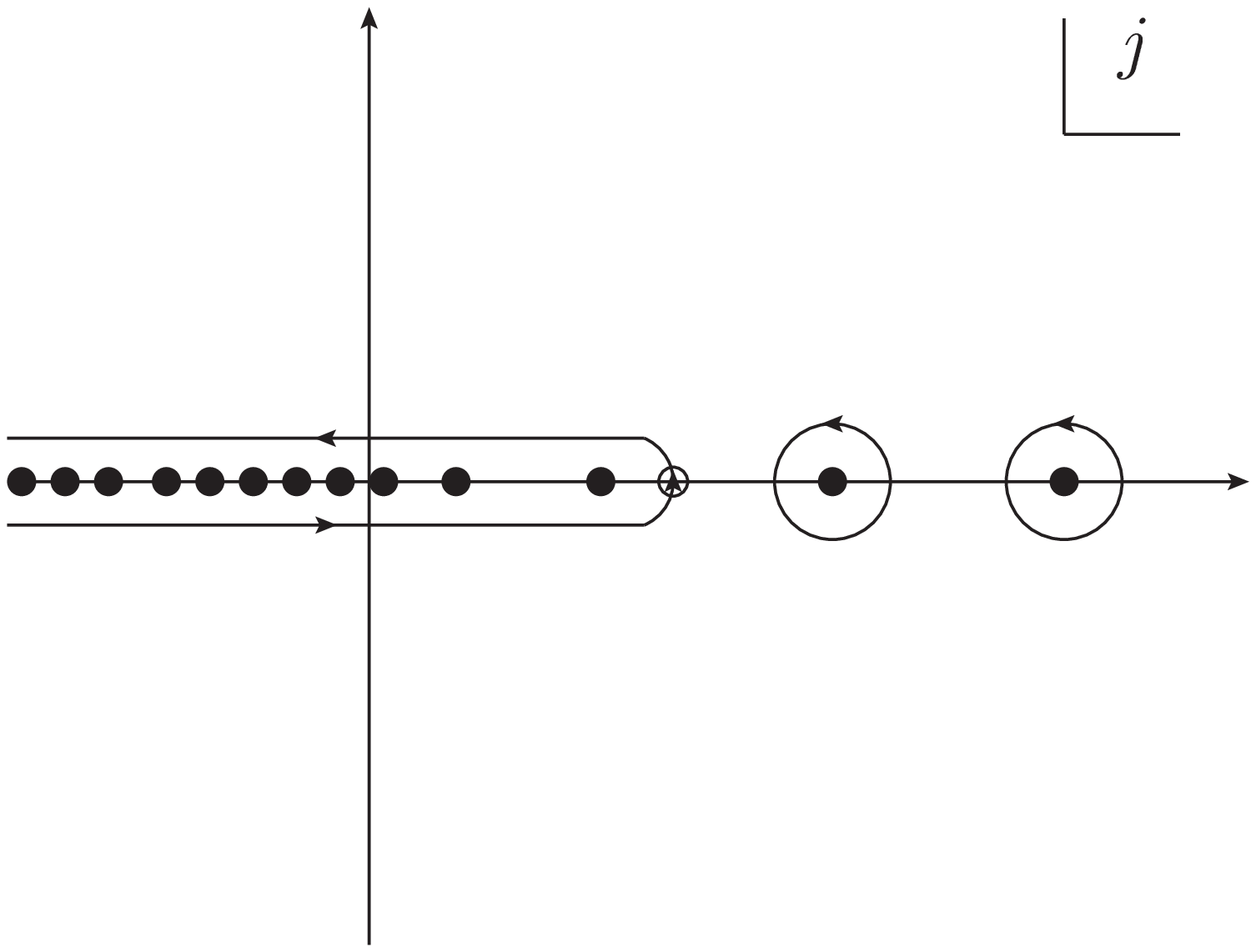} \\  
  (a) &  (b) & (c) & (d) 
 \end{tabular}
\caption{\label{fig:a} 
 Singularities and integration contours in the complex $j$-plane. 
 The hard-wall model is assumed for (a) negative $t$ and (b) sufficiently 
 large positive $t$, while a holographic model for asymptotic free 
 running coupling is assumed in (c) with a smaller $t$, and in (d) 
 with a larger $t$.
 Black dots are poles, wiggling lines in (a, b) are branch cuts, and 
 open circles in (a--d) denote saddle points of the amplitude on the 
 complex $j$-plane.
}
  \end{center}
\end{figure}

Before discussing
kinematical parameter ($x,t,q^2$) dependence of the DDVCS amplitude
in gravity dual, let us clarify a couple of conceptual issues associated with 
Pomerons. Using the explicit form of the Pomeron 
kernel (\ref{eq:pomeron kernel}) and 
Pomeron wavefunctions (\ref{psi}), amplitudes $I_i$ ($i = 0,1$) in 
(\ref{eq:Ii-def}) can be rewritten (see \cite{NW} for details) as 
\begin{align}
 I_i (x, \eta = 0, t, q^2) & \simeq
  \sqrt{\lambda} \int^\infty_{-\infty} d\nu 
  \left[ - \frac{1 + e^{-\pi i j_\nu}}{ \sin \pi j_\nu}\right]
  \frac{1}{\Gamma^2(j_\nu/2)} \; 
  \left[C^{(i)}(j,q)\right]_\mu \; \left[A_{hh}\right]_\mu, 
  \label{eq:Im-Ii-cm} 
\end{align}
where 
\begin{eqnarray}
 \left[C^{(i)}(j,q)\right]_\mu & = & \left[\frac{1}{R^3} \int dz \sqrt{-g(z)} 
   P^{(i)}_{\gamma^* \gamma^*}(z) e^{-2A(z)} 
   \left(\frac{z}{R}\right)^{i\nu} (R z)^{j_\nu} \right] 
  \times (R \mu)^{i\nu - j_\nu}, 
 \label{eq:C(i)-def} \\
 \left[A_{hh}\right]_\mu & \simeq  &  \frac{1}{(R\mu)^{i\nu-j_\nu}} \times 
 \left[
 \frac{c'_s}{\kappa_5^2} \int dz' \sqrt{-g(z')} P_{hh}(z')
   \left[ \frac{e^{-2A(z')}W^2}{4\sqrt{\lambda}}\right]^{j_\nu} \right.
   \nonumber \\
  & & \qquad \left.
   \left[\frac{e^{(j_\nu - 2)A(z')}}{K_{i\nu}(\sqrt{-t} R)} 
      \left( K_{i\nu}(\sqrt{-t}z')
             - \frac{K_{i\nu}(\sqrt{-t}/\Lambda)}
                    {I_{i\nu}(\sqrt{-t}/\Lambda)}
               I_{i\nu}(\sqrt{-t}z')\right)\right]
   \right]; 
\label{eq:Ahh-def}
\end{eqnarray}
a parameter $\mu$ of mass dimension $+1$ is introduced in
(\ref{eq:C(i)-def}, \ref{eq:Ahh-def}) in a way the observables $I_i$ 
are unaffected. One can change the integration variable of
(\ref{eq:Im-Ii-cm}) from $\nu$ to $j = j_\nu$; now the amplitudes 
$I_i$ are given by integration over the complex $j$-plane, and the
contour becomes the one in Figure~\ref{fig:a}~(a).

The factor $\left[C^{(i)}(j,q)\right]_\mu$ is now regarded as a function of $j$, 
and also depends on $q^2$ and $\mu$, but not on $t$ or $x$.
Its asymptotic form for $q^2\gg \Lambda^2$ is given by
\begin{equation}
 \left[C^{(i)}(j,q)\right]_\mu \simeq c_J^2 
   \left(\frac{\mu}{q}\right)^{\gamma(j)} 
   \frac{1}{(q^2)^{j}} \bar{c}^{(i)}_{i\nu_j},
\end{equation}
with a dimensionless constant of order unity $\bar{c}^{(i)}_{i\nu_j}$ 
that depends only on $j$. Here, $\gamma(j) \equiv i\nu_j - j$, and 
$\nu_j = \nu(j)$ is the inverse function of $j = j_\nu$ (\ref{eq:j-jnu}).
$x$ dependence and $t$ dependence of the amplitudes $I_i$ come from the 
other factor $\left[A_{hh}\right]_\mu$.
It can be rewritten as
\begin{align}
 \left[A_{hh}\right]_\mu \simeq  
  c'_s \left(\frac{W^2}{4\sqrt{\lambda}}\right)^{j}
  \left(\frac{\Lambda}{\mu} \right)^{\gamma(j)} 
  g^h_{i\nu_j}(\sqrt{-t}/\Lambda)
   \simeq 
  c'_s
  \left(\frac{1}{4\sqrt{\lambda}x}\right)^{j} 
  \left(q^2 \right)^{j}
  \left( \frac{\Lambda}{\mu} \right)^{\gamma(j)} 
  g^h_{i\nu_j}(\sqrt{-t}/\Lambda) \label{eq:Ahh--gh},
\end{align}
where $g^h_{i\nu_j}(\sqrt{-t}/\Lambda)$ is a dimensionless function of $j$ and $(\sqrt{-t}/\Lambda)$.
For the final expression, we used $W^2\simeq q^2/x$ which holds at small $x$.
Combining both, one finds that 
\begin{align}
I_i & \simeq
   c'_s \sqrt{\lambda} \int^{-\infty+i\epsilon}_{-\infty-i\epsilon} d j
  \frac{|\partial \nu_j/\partial j| \, c_J^2}{\Gamma^2(j/2)} 
  \left[- \frac{1+e^{-\pi i j}}{\sin (\pi j)}\right]
  \sfrac{1}{4\sqrt{\lambda}x}^{j}\sfrac{\Lambda}{q}^{\gamma(j)}
  \bar{c}^{(i)}_{i\nu_j} \:g^h_{i\nu_j}(\sqrt{-t}/\Lambda).
\label{eq:Im-Ii}
\end{align}
This is in the form of inverse Mellin transformation, and the
integration variable $j$ is identified with the complex angular
momentum (complex spin).\footnote{Since we restrict ourselves to the scattering at 
$\eta=0$, total derivative operators in field-theory language do not 
contribute to the OPE of the scattering amplitude. Thus, there is no 
subtleties in what this $j$ is here.} 

Now, physical meaning of the separation between $\left[C^{(i)}_{i\nu_j}\right]_\mu$
and $\left[A_{hh}\right]_\mu$ (or $g^h_{i\nu_j}(\sqrt{-t/\Lambda})$) is 
clear. By changing the integration contour in the $j$-plane, 
(\ref{eq:Im-Ii-cm}, \ref{eq:Im-Ii}) can be rewritten as 
\begin{equation}
\label{eq:Ii-in-jsum}
 I_i \simeq \sum_{j \in 2\N} 
  \frac{ 4\sqrt{\lambda} \left| \frac{\partial i\nu_j}{\partial j}
	\right|}{\Gamma^2(j/2)}
  \left[c_J^2  \left(\frac{\mu}{q}\right)^{\gamma(j)} 
   \frac{1}{(q^2)^{j}} \bar{c}^{(i)}_{i\nu_j} \right] 
  c'_s \left(\frac{2 q \cdot p}{4\sqrt{\lambda}}\right)^j
  \left[\left(\frac{\Lambda}{\mu} \right)^{\gamma(j)} 
  g^h_{i\nu_j}(\sqrt{-t}/\Lambda) \right]. 
\end{equation}
This is regarded as an OPE form of $I_i$. The first factor 
in $\left[ \cdots \right]$, which comes from
$\left[C^{(i)}(j,q)\right]_\mu$, is regarded as the Wilson coefficient of 
OPE for a spin $j \in 2\N$ operator; the parameter $\mu$ is now
identified with the renormalization scale, because of its appropriate scaling behavior 
determined by the anomalous dimension $\gamma(j)$ of the ``twist-2''
spin $j$ operator. The second factor in $\left[ \cdots \right]$ is
identified with the spin $j$ form factor, which is
the coefficient of the $[p^{\mu_1} \cdots p^{\mu_j}]$ 
term of the hadron matrix element of the spin $j$ operator renormalized 
at the scale $\mu$. The gravity dual expression (\ref{eq:Ahh-def})  
justifies such an interpretation \cite{GKP}. 

Knowing physical meaning of these factors in the {\it scattering
amplitude} (\ref{eq:Im-Ii-cm}) in a gravity dual model, one can define 
a {\it GPD} even in the model, which corresponds to a strongly coupled 
gauge theory. GPD as a function of $x$ and $t$ (we only consider the 
$\eta = 0$ case in this letter) is defined as an inverse Mellin
transform of form factors of twist-2 
spin $j$ operators 
(the second factor in [$\cdots$] of (\ref{eq:Ii-in-jsum})).
 The scattering amplitude $I_i$ is given by
convolution of this GPD, inverse Mellin transform of the Wilson
coefficient and that of the signature factor 
$-[1+e^{-\pi i j}]/\sin (\pi j)$, just like in perturbative QCD 
factorization formula. The inverse Mellin transform of the signature 
factor gives rise to a light-cone singularity of a propagating parton 
(like the one in \cite{Ji-PRD}), even in the gravity dual description. 
The GPD determined in this way is essentially\footnote{Since GPD 
is defined as the inverse Mellin transform of form factors of 
``twist-2'' spin $j$ operators, it would become different when the
normalization of the operators were changed in a $j$-dependent
manner. We do not pay such a careful attention in this article. 
We claim similarity between ${\rm Im} \; I_i$ and GPD after replacement of $q^2$
by $\mu^2$ only at this level of precision. } the same as 
${\rm Im} \; I_i$, with $q^2$ of ${\rm Im} \; I_i$ replaced by the
renormalization scale $\mu^2$; thus, various statements on 
${\rm Im} \; I_i$ in the rest of this section are also applied to 
the GPD after $q^2$ is replaced by $\mu^2$.

Now that the field theory OPE interpretation of the gravity dual amplitude
(\ref{eq:Im-Ii-cm}) is clarified, let us go back to the 
amplitude (\ref{eq:Im-Ii-cm}) and explicit expressions 
(\ref{eq:C(i)-def}, \ref{eq:Ahh-def}) once again. We will now clarify 
how this string theory amplitude on a warped background is related 
to the traditional Regge phenomenology ansatz.
It should be noted that the DDVCS amplitudes $I_i$ in gravity dual 
(\ref{eq:Im-Ii}) do not have a Pomeron pole like $1/(j-\alpha(t))$ 
in their $j$-plane representation apparently. There was once a pole 
$1/(j-j_\nu)$ at the stage of (\ref{eq:pomeron kernel}), but it is gone 
in (\ref{eq:Im-Ii}), after picking the residue to evaluate an integral 
in (\ref{eq:pomeron kernel}). Nevertheless, one can see that the expression
(\ref{eq:Im-Ii}) may have, in fact,  many poles in the $j$-plane,
rather than a single pole or none. 

To see this, we can use 
Kneser--Sommerfeld expansion of Bessel functions in the hard wall model
to rewrite $[\Gamma_{hh \P^*}(j,t)]_\mu \equiv (\Lambda/\mu)^{\gamma(j)} 
g^h_{i\nu_j}(\sqrt{-t}/\Lambda)$ as \cite{NW}
\begin{eqnarray}
 \left[\Gamma_{hh \P^*}(j, t)\right]_\mu
  &= & \sum_{n=1}^{\infty} \frac{-2}{t-m^2_{j,n}}
      \frac{\gamma_{hh \P n}(j)}{\Lambda^{j-2}}
      \frac{\left(\frac{m_{j,n}}{2}\right)^j 
      \left(\frac{m_{j,n}}{2 \mu}\right)^{\gamma(j)}}
           {[J'_{i\nu_j}(j_{i\nu_j,n})]}
      \frac{2}{\Gamma(i\nu_j)}
      \left[\frac{R^3}{\kappa_5^2}\right]^{1/2},
\label{eq:Pomeron-form-factor-3pt-decay} \\
 \gamma_{hh \P n}(j) & = & \frac{1}{\kappa_5^2} \int dz \sqrt{-g} \; 
   P_{hh}(z) \; e^{- 2 j A} \; \psi^{(j)}_n(z) \times
   (R \Lambda)^{j}.
\label{eq:P-h-h-3pt-coupling}
\end{eqnarray}
The Pomeron trajectory (that contains graviton) of the Type IIB string
theory on 10-dimensions (or on $AdS_5$ after dimensional reduction on
$W$) gives rise to a Kaluza--Klein tower of infinitely many Pomeron 
trajectories in hadron scattering on 3+1 dimensions. 
These trajectories are labeled by the Kaluza--Klein excitation level
$n$; the masses of spin $j \in 2\N$ hadrons are $m_{j,n}$, and their 
wavefunctions on $AdS_5$ are $\psi^{(j)}_n(z)$. 
The factor $1/(t-m_{j,n}^2)$ in (\ref{eq:Pomeron-form-factor-3pt-decay}) 
becomes a $t$ dependent pole in the $j$-plane, the Pomeron pole, for any 
one of $n$'s. In the hard wall model, the Pomeron trajectory 
($j=\alpha_{\P,n}(t)$ relation set by $t=m_{j,n}^2$) and the Pomeron 
wavefunction for the $n$-th trajectory are obtained holomorphically in $j$ 
(not just for $j \in 2\N$) as in 
\begin{equation}
 m_{j,n} = \Lambda j_{i\nu_j,n}, \qquad 
\label{eq:psi-tilde:discrete-eigenfunction-of-deltaj}
  \psi^{(j)}_n(z) = e^{(j-2) A}
      \frac{J_{i\nu_j}(m_{j,n} z)}{J'_{i\nu_j}(j_{i\nu_j,n})}
      \left[\frac{\kappa_5^2}{R^3}\right]^{1/2},
\end{equation}
where, $j_{\mu,n}$ is the $n$-th zero of Bessel function $J_\mu(z)$. 
$m_{j,n}$ is read out from the denominator in (\ref{eq:Ahh-def}); the
wavefunction $\psi^{(j)}_n(z)$ satisfies an equation of motion of 
a spin $j$ field on $AdS_5$, just like (\ref{psi}) does. 
Although explicit expressions above rely heavily on the hard wall model, 
conceptual understanding itself is quite general, and is applicable 
at least to any asymptotically conformal gravity dual models.  

Therefore, gravity dual descriptions of strongly coupled gauge theories  
come up with a following picture of Pomeron exchange amplitude. 
Individual Pomerons in the Kaluza--Klein tower couple to the target 
hadron with a coupling $\gamma_{hh\P n}(j)$ in
(\ref{eq:P-h-h-3pt-coupling}), which do not show any power-law fall-off 
behavior in large negative $t$. Only after all the Pomeron couplings 
$\gamma_{hh\P n}(j)$ and Pomeron propagators 
$1/(t-m_{j,n}^2) \propto 1/(j-\alpha_{\P,n}(t))$ are combined as 
in (\ref{eq:Pomeron-form-factor-3pt-decay}), do we obtain what 
we might call a ``Pomeron form factor'' $[\Gamma(j,t)]_\mu$, which has 
a power-law behavior in $\sqrt{-t}$ (see (\ref{eq:g-h-tilde})).
Such a relation between a form factor of a conserved current and 
a combination of a Kaluza--Klein tower of hadrons, three point 
couplings and decay constants has been known for fixed spins 
(such as $j=1$ and $j=2$) \cite{Strassler-FF}. 
The relation (\ref{eq:Pomeron-form-factor-3pt-decay}) is regarded 
as an analytic continuation in $j$ of the one for graviton (spin $j=2$). 

\subsection{Saddle point in the $j$-plane}
\label{ssec:saddle}

It was a conventional wisdom of traditional Regge phenomenology 
that behavior of hadron scattering amplitudes at high energy 
are governed by the position of singularities in the complex $j$-plane. 
The same is true in gravity dual description of strongly coupled gauge
theories.
 Singularities in scattering amplitudes in the complex $j$-plane representation depend on choice of gravity dual models.
In case of the hadron--virtual ``photon'' scattering,
however,
the scattering amplitude can be approximated
at saddle point in the $j$-plane
(within a certain kinematical region 
which we call ``saddle point phase'' in \S\ref{ssec:PT}).
In this case,
the expression of the amplitude becomes
not directly dependent on the singularities, and hence,
 detail of gravitational dual is irrelevent.
In this subsection, we employ the hard wall model,
and study the behavior of this scattering amplitude.

In the hard wall model, there are no isolated poles in the complex $j$-plane for negative $t$---physical kinematics---except the branch 
cut that extends to negative $j$ along the real axis, 
Figure \ref{fig:a}(a); $(t-m_{j,n}^2)$ never vanishes for $t < 0$.
Thus, the $j$ integral of (\ref{eq:Im-Ii}) along the contour 
in Figure~\ref{fig:a}~(a) is evaluated by the saddle point method  
for small $x$ \cite{Hatta:2007he}.
For $0 \leq -t \lesssim \Lambda^2$, the saddle point 
value of $j$ is given by\footnote{
The integrand of the Pomeron kernel (\ref{eq:pomeron kernel}) is
 reliable at $|j|\sim {\cal O}(1)$, but not at $|j|\gtrsim \sqrt{\lambda}$
\cite{Gubser:2002tv}.
Therefore, we note that kinematical variables ($x$, $q^2$ and $t$)
 are required to be consistent with $j^\ast\sim {\cal O}(1)$.
}
\begin{equation}
 j^* = j_{\nu^*}, \qquad i\nu^* = 
  \frac{\sqrt{\lambda}
  \ln(q/\Lambda)}{\ln\left[(q/\Lambda)/(\sqrt{\lambda}x)\right]}, 
\label{eq:nu ast for small -t}
\end{equation}
and 
\begin{equation}
 {\rm Im} \; I_i(x,\eta=0, t , q^2)\sim 
    \sfrac{1}{\sqrt{\lambda}x}^{j^*}
    \sfrac{\Lambda}{q}^{\gamma(j^*)}
    g^{h}_{i\nu_{j^*}}\left(\sqrt{-t}/\Lambda \right); 
\label{eq:J for small -t}
\end{equation}
the form factor $g^h_{i\nu}(\sqrt{-t}/\Lambda)$ has a
dimensionless non-zero limit of order unity when $-t \rightarrow 0$;  
it begins to fall off in power-law\footnote{Thus, for $\Lambda^2 \ll -t$,
 the saddle point becomes 
\begin{equation}
  i\nu^\ast \left( q/\Lambda, x, -t \gg \Lambda^2 \right)=
   \sqrt{\lambda}\frac{\ln(q/\sqrt{-t})}
                      {\ln\sfrac{q/\sqrt{-t}}{\sqrt{\lambda}x}}.
 \label{eq:nu ast for large -t}
\end{equation}
} as 
\begin{align}
 g^h_{i\nu}(\sqrt{-t}/\Lambda) \simeq 
 \sfrac{\Lambda}{\sqrt{-t}}^{-\gamma(j)+2\Delta-2} \tilde g^h_{i\nu_j}
\label{eq:g-h-tilde}
\end{align}
for larger momentum transfer $\Lambda^2 \ll -t$. 
Here, $\Delta$ is the scaling dimension of the scalar field on $AdS_5$
containing the target hadron $h$, and $\tilde g^h_{i\nu}$ is 
a $t$-independent (but $\nu_j$-dependent) constant of order unity.
Note, in particular, that the $t$-dependence of the scattering amplitude 
is given by the form factor that is once analytically continued to
complex $j$-plane and then evaluated at the saddle point. 
The Regge factor $(W^2)^j$ of string theory amplitude justifies focusing
on a small range of $j$ (or $\nu$) around the saddle point value at
high-energy scattering; the power-law behavior in $\sqrt{-t}$
follows from the power-law wavefunction of the target hadron $P_{hh}(z')$ and 
exponential cut-off of the Pomeron wavefunction, 
$K_{i\nu}(\sqrt{-t}z)$ in (\ref{eq:Ahh-def}) in particular, in the
limited range of $\nu_j$.

The saddle point method provides a good approximation to the scattering 
amplitude for $\ln(1/x)/\sqrt{\lambda} \gg 1$. It should be noted, 
however, that it allows us to keep all-order contributions 
in $i\nu^\ast=\sqrt{\lambda} \ln(q/\Lambda)/\ln(1/x)$,
which is not necessarily small and can be as large as ${\cal O}(\lambda^{1/4})$.
Thus, amplitudes and observables are expressed as functions of $i\nu^\ast$ (or $j^\ast$).
This makes easy to understand their dependence of kinematical variables ($x,t,q^2$).

Equation (\ref{eq:J for small -t}) 
clearly shows the importance of the value of the saddle point of the 
$j$-plane amplitude. 
To see this more explicitly, let us define 
\begin{align}
 \gamma_\text{eff}(x,t,q^2)&=\frac{\partial \ln
 [x \; I_i(x,\eta=0,t,q^2)]}{\partial \ln (\Lambda/q)}, \qquad &
 \lambda_\text{eff}(x,t,q^2)&=\frac{\partial \ln [x
 I_i(x,\eta=0,t,q^2)]}{\partial \ln (1/x)}.
\label{eq:def-exponents}
\end{align}
It is straightforward to see that they are given by
\begin{align}
 \gamma_\text{eff}(x,t,q^2)&=\gamma(j^*), & \qquad 
 \lambda_\text{eff}(x,t,q^2)&=j^* -1.
\end{align}
These effective exponents $\gamma_{\rm eff.}$ and $\lambda_{\rm eff.}$ 
depend on kinematical variables $x, q^2$ and $t$ only through the 
saddle point value $j^*$. The ratio 
$\rho = {\rm Re} \; I_i/{\rm Im} \; I_i=\tan\left(\frac{\pi}{2}(j^\ast-1)\right)$ is also 
related directly to the saddle point value $j^*$. 

We can see 
from (\ref{eq:nu ast for small -t}, \ref{eq:nu ast for large -t})
that $j^*$ becomes large for large $q^2$ and small for small $x$.
Thus, at a given renormalization scale $\mu^2$ 
(replace $q^2$ in ${\rm Im} \; I_i$), GPD in gravity dual still 
increases in the DGLAP evolution (that is, $\gamma_{\rm eff} < 0$) 
for small enough $x$ such that the saddle point value $j^*$ is still less than 2.
Even at such a small value of $x$, however, GPD eventually begins to decrease
(that is, $\gamma_{\rm eff}$ becomes positive) for large enough $\mu^2$.
Such a behavior of GPD---qualitatively the same as in the real 
world QCD---in the DGLAP evolution was anticipated 
in \cite{PolchinskiJHEP0305:0122003}; this is indeed realized for finite $\mu^2$ 
in gravity dual, when both $q^2$ $(\mu^2)$ and $x$ dependence are included in the saddle point approximation.
The other parameter $\lambda_{\rm eff}$ characterizing 
the $x$ evolution is known to increase gradually for larger $q^2$ in the 
real world QCD \cite{lambda-rise}. As already seen in
\cite{Brower:2010wf}, it does follow from gravity dual as well; 
we understand that this phenomenon is also essentially due to the increase 
of the saddle point value $j^*$ for larger $q^2$.
The same behavior is also obtained in perturbative QCD (See \cite{DLLA}). 

\begin{figure}[tbp]
  \begin{center}
  \includegraphics[scale=0.7]{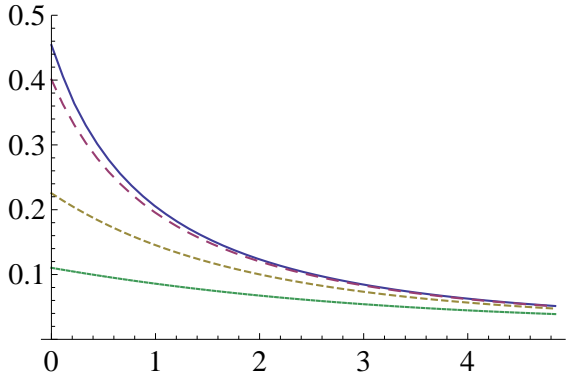} 
\caption{\label{fig:slope} (color online) 
Slope $B$ of the forward peak in DDVCS. 
The dimensionless value $B \times \Lambda^2$ is shown as a function 
of $i\nu^*$ (\ref{eq:nu ast for small -t}); from top to bottom, blue 
(solid line) curve is for $\sqrt{-t}/\Lambda \simeq 0.01\mbox{--}0.1$, 
red (long dashed) one for $\sqrt{-t}/\Lambda = 1.$, yellow (dashed) one 
for $\sqrt{-t}/\Lambda = 3.$, and green (short dashed) one for 
$\sqrt{-t}/\Lambda = 6$. 
For more details, see \cite{NW}.}
  \end{center}
\end{figure}

The $t$ dependence of the scattering amplitude is characterized by 
the slope parameter of the forward peak (also known as $t$-slope parameter), which we define for non-skewed 
scattering as 
\begin{align}\label{eq:def-slope-parameter}
B_i(x, \eta=0, t, q^2)= 2 \;  \frac{\partial}{\partial t}
  \ln {\rm Im} \; I_i(x,\eta=0,t,q^2).
\end{align}
The $t$-dependence (and hence the slope parameter) comes entirely 
from the form factor for the physical kinematical region $t \leq 0$ 
in the hard wall model. The $t$-slope parameter at $t=0$ in such a case 
can be regarded as the charge radius square of the hadron under 
``spin-$j^*$ probe''. 
Explicit expressions for the form factor in the hard wall model allow us 
to calculate the $t$-slope parameter; see Figure~\ref{fig:slope}.
The larger the spin $j^*$ (and hence $i\nu^*$), the smaller the slope.
Therefore, through (\ref{eq:nu ast for small -t}), the slope parameter 
decreases for larger $q^2$, a prediction of a gravity dual model which 
cannot be made within perturbative QCD.
 
We can also see from 
(\ref{eq:nu ast for small -t}, \ref{eq:nu ast for large -t})
that the saddle point value $j^*$ depends weakly on $\ln (1/x)$ or 
$\ln (W^2/\Lambda^2)$ than on $\ln(q^2/\Lambda^2)$ for small $x$, and 
the dependence is in the opposite direction.
Thus, the $\ln(1/x)$ dependence (or $\ln(W^2/\Lambda^2)$ dependence)
of the slope parameter $B$ must be weaker than its $\ln(q^2/\Lambda^2)$
dependence. This property of $B$, shared by $\gamma_{\rm eff}$,
$\lambda_{\rm eff}$ and $\rho$, is an immediate consequence of the 
fact that the scattering amplitude is well approximated by the saddle
point method on the $j$-plane integral. This is a fairly robust feature 
of the saddle point approximation, and does not rely on specific details 
of the hard wall model. 

The saddle point turns out to be an important concept also in 
the scattering amplitude ${\rm Im} \; I_i$ in the impact parameter
space, which is obtained by taking a Fourier transform in the transverse 
direction of the momentum transfer $(p_1-p_2)$. The $x$-dependent 
parton density profile in the transverse direction obtained in this 
way \cite{b-space} in gravity dual shows Gaussian profile at large 
impact parameter $b$, but is larger than the simple Gaussian form for 
smaller $b$ (\cite{NW}; see also \cite{Brower:2010wf}; deviation from
the simple Gaussian profile is an immediate consequence of the fact that
the 4D leading trajectory $j=\alpha_{\P, n=1}(t)$ is not perfectly linear). 
This core of larger parton density has approximately a linear exponential
profile, $e^{-m_{j^*,1} b}$.
The effective mass scale $m_{j^*,n=1} = m^{(\nu^*)}_{n=1}$ gradually
changes as a function of $b$, and the linear exponential form smoothly 
turns into the Gaussian form for larger $b$, when $i\nu^*$ becomes of 
order unity. See \cite{NW} for more. 

\subsection{Pole--Saddle Point Crossover}
\label{ssec:PT}
Although we saw that
the saddle point method well approximated the DDVCS amplitude
for physical kinematical region $t\le 0$ in the hard-wall model,
it does not in general.
Even in small $x$, whether or not the scattering amplitude is well approximated 
by the saddle point method, depends on singularities of the amplitude 
in the $j$-plane representation, and hence on the gravity dual model
one considers, and also the values of the kinematical variables 
$x, q^2$ and $t$. Although all the asymptotically conformal gravity dual
models have a branch cut that is stretched to large negative $j$, there
may also be some isolated poles in the $j$-plane as in Figure~\ref{fig:a}~(b).
The hard wall model does not have such a pole for physical kinematical
region $t \leq 0$ (there are for sufficiently positive $t$), but there 
may be some for other UV conformal models that have different (and
faithful to string theory construction) infrared geometry. Even more
interesting are gravity dual models that are asymptotically free, where 
the cut is replaced by isolated singularities (Figure~\ref{fig:a}~(c,
d)) \cite{BrowerJHEP0712:0052007}.

When the saddle point (open circle in the figure) has a larger real part 
than any one of the singularities in the complex $j$-plane, then the 
integration contour in the $j$-plane should simply be chosen so that it passes 
through the saddle point, as in Figure~\ref{fig:a}~(a, c). When some of the
singularities have larger real parts than the saddle point value $j^*$, 
however, it is more convenient to take the contour as 
in Figure~\ref{fig:a}~(b, d), so that the scattering amplitude is given
by contributions from finite number of isolated Pomeron poles
$j = \alpha_n(t)$ ($n=1,2,\cdots$) and by a continuous integration over 
a contour passing through 
the saddle point. We refer to the two situations 
as  saddle point phase and leading pole phase (or leading singularity phase),
 respectively. 
Such observables as $\lambda_{\rm eff}$, $\gamma_{\rm eff}$, $\rho$ and $B$ exhibit totally 
different dependence on the kinematical variables $x, q^2$ and $t$ in
the two phases. 
In a given theory (i.e., in a given gravity dual model), 
one always 
enters into the saddle point phase for sufficiently large $q$ or
sufficiently negative $t$.
In asymptotically free theories, it is likely  
that the leading singularity phase also exists for sufficiently small $x$ and 
not so large negative $t$, even in the physical kinematical region $t\leq 0$.

The transition between the two phases is not singular but 
is a (smooth) crossover for finite $x$. 
This is because the saddle point approximation is never exact, and 
the ``saddle point'' should be thought of as a sort of diffuse object
for finite $x$. Subleading singularities may also give rise to
significant corrections to the amplitude simply given by the leading 
pole $j=\alpha_1(t)$ for finite $x$, too. 
The transition becomes a singular phase transition only in the 
extreme small $x$ limit.

\section{Lessons to Learn}

It is true that gravity dual calculation employs a 
background that corresponds to large 't Hooft coupling even 
at energy scale much larger than the hadronic scale $\Lambda$. 
Still, there are surprisingly many qualitative features in the gravity 
dual hadron--virtual ``photon'' scattering amplitude that are
in common with the scattering amplitude in the real world QCD.
Scattering amplitudes in gravity dual 
have $\ln (q/\Lambda)$ and $\ln (1/x)$ scaling governed by
$\gamma_{\rm eff} = \gamma(j^*)$ and $\lambda_{\rm eff} = j^* -1$
in the saddle point phase,
and this is the same qualitatively as the prediction of the saddle point method
 in perturbative QCD,
 as we have already seen in 
\S \ref{ssec:saddle}. 
The only difference between gravity dual and real world QCD is in the choice of 
anomalous dimension, $\gamma(j)$. Qualitative features are shared by 
both, and are controlled by the saddle point value $j^*$. 

Qualitative features in $t$-dependence also show agreements.
The gravity dual amplitude continues 
to the power-law fall-off behavior at large momentum transfer 
$\Lambda^2 \ll -t$. This property, which is expected to hold in the 
real world QCD theoretically \cite{Lepage:1980fj} and confirmed 
experimentally, was difficult to be consistent with the traditional Regge 
phenomenology, but this problem is now overcome in gravity dual 
on {\it warped} spacetime (cf. \cite{Polchinski2002, PolchinskiJHEP0305:0122003}).
Moreover,
the $t$-slope parameter of (\ref{eq:def-slope-parameter}) and its result 
in Figure~\ref{fig:slope} for $\eta = 0$ in gravity dual at saddle point phase
nicely agrees with that in DVCS differential cross section \cite{DVCS-slope-on-q}, 
in that the slope parameter $B$ decreases for larger $\ln (q/\Lambda)$, 
and is less
sensitive to $\ln (1/x)$ or $\ln (W/\Lambda)$. 
Such observation suggests the (analytically continued) spin $j$
form factors $[\Gamma_{hh \P^*}(j,t)]_\mu$ in both a gravity dual model
and the real QCD are similar to each other.

With so many basic qualitative features that gravity dual shares with 
the real world QCD, it is thus tempting to try to extract some lessons 
from the hadron--virtual ``photon'' amplitude in gravity dual.
The origin of such similarity at the qualitative level becomes clear 
in the complex $j$-plane representation,
where GPD is given by inverse Mellin transformation:
\begin{equation}
 H(x,\eta=0, t; \mu^2) \sim \int \frac{dj}{2\pi i} 
  \left(\frac{1}{x}\right)^{j}
\left[  \left(\frac{\Lambda}{\mu}\right)^{\gamma(j)}
  g^{h}_{i\nu_j}(\sqrt{-t}/\Lambda)\right].
\label{eq:GPD-in-j}
\end{equation}
Indeed, it is always possible to describe scattering amplitude
 by the $j$-plane integral
 in any theories,
independent of whether the scattering is based on the real QCD or on the strongly coupled gauge theory
studied in gravity dual;
this is because Mellin transformation is only a mathematical transformation.
This $j$-plane integral also comes form OPE,
notion of which is well-defined even in strongly coupled theories \cite{PolchinskiJHEP0305:0122003}.
GPD in the $j$-plane representation (\ref{eq:GPD-in-j})
is given by dropping the Wilson coefficient of OPE from the scattering amplitude 
${\rm Im}\: I_i$ (\ref{eq:Im-Ii-cm}),
so the spin $j$ form factor 
(reduced matrix element of twist-2 spin $j$ operator),
 which is the content of [$\cdots$] in (\ref{eq:GPD-in-j}),
determines GPD.
The spin $j$ form factor is decomposed into two parts:
RG evolution part $(\Lambda/\mu)^{\gamma(j)}$, and
 form factor at renormalization scale $\mu=\Lambda$,
 $g^{h}_{i\nu_j}(\sqrt{-t}/\Lambda)$;
both show common properties in the real QCD and in gravity dual.
The anomalous dimensions of the twist-2 spin $j$ operator $\gamma(j)$ in both theories 
are qualitatively similar \cite{BrowerJHEP0712:0052007},
and $g^{h}_{i\nu_j}(\sqrt{-t}/\Lambda)$ has the 
power-law fall-off behavior at $-t\gg\Lambda^2$ in common.

The behavior of GPD is determined by the saddle point, or alternatively,
by the leading singularity
depending on which phase a set of parameters $(x,t,\mu^2)$ sits in.
This classification is applicable in any theories, not just in gravity dual.
Then it is important to know which phase a given set $(x,t,\mu^2)$ sits in. 
As we have pointed out, the behaviors of 
$\gamma_\text{eff}$ and $\lambda_\text{eff}$ 
observed in HERA for DIS \cite{lambda-rise}
are successfully explained by the predictions of the saddle point phase
and are inconsistent with the prediction of the leading pole phase
(or the leading singularity phase)
\cite{NW}.
Therefore,
it is very likely that 
the (most of) kinematical region of DVCS
that has been explored in HERA measurements 
is in the saddle point phase,\footnote{
In the standard parametrization of DVCS cross section
$ \frac{d \sigma_{\rm DVCS}(\gamma^*p \rightarrow \gamma p)}{dt} \sim 
 \frac{\alpha_{QED}^2}{\Lambda^{4}} \times
 \left(\frac{W}{\Lambda}\right)^{\delta}
 \left(\frac{\Lambda^2}{q^2}\right)^n$,
the parameters $(\delta, n)$ are given by
$\delta=4(j^\ast-1)$ and $n=\gamma(j^\ast)+2j^\ast$
in the saddle point phase.
Thus, the saddle point phase implies rise of $\delta$ for larger $q^2$.
HERA measurement \cite{delta-hera} gives
$\delta=0.44\pm0.19$ for $q^2=2.4$ GeV$^2$,
$\delta=0.52\pm0.09$ for $q^2=3.2$ GeV$^2$,
$\delta=0.75\pm0.17$ for $q^2=6.2$ GeV$^2$,
$\delta=0.84\pm0.18$ for $q^2=9.9$ GeV$^2$, and
$\delta=0.76\pm0.22$ for $q^2=18$  GeV$^2$
in ZEUS,
and
$\delta=0.61\pm0.10\pm0.15$ for $q^2=8$ GeV$^2$,
$\delta=0.61\pm0.13\pm0.15$ for $q^2=15.5$ GeV$^2$, and
$\delta=0.90\pm0.36\pm0.27$ for $q^2=25$ GeV$^2$,
in H1.} and GPD is approximately
 given by
\begin{align}
  H(x,\eta=0,t,\mu^2)\sim
  \left(\frac{1}{x}\right)^{j^*}
  \left(\frac{\Lambda}{\mu}\right)^{\gamma(j^*)}
  g^{h}_{i\nu_{j^*}}(\sqrt{-t}/\Lambda).   
\label{eq:GPD-saddle-approx}
\end{align}
Most of the observed properties of the $t$-slope parameter $B$ of DVCS in HERA \cite{DVCS-slope-on-q}
can be understood only from the fact that the kinematical region 
is in the saddle point phase (see \cite{NW}).
A GPD model with a specific choice of $g^{h}_{i\nu_{j^\ast}}(\sqrt{-t}/\Lambda)$
in \cite{Mueller:2006pm} belongs to this category.\footnote{Reference
 \cite{Mueller:2006pm} introduces an ansatz $g^h_{i\nu_j}\sim (j-\alpha(t))^{-1}(1-t/\Lambda^2)^{-p}$,
inspired by a leading Pomeron pole $(j-\alpha(t))^{-1}$ and a power-law fall-off for $(-t)\gg\Lambda^2$.
Our result (\ref{eq:Pomeron-form-factor-3pt-decay}) is conceptually different
 from this model;
 each Pomeron pole term with a Kaluza--Klein excitation level $n$
does not show the behavior of power-law fall-off,
but the power-law (\ref{eq:g-h-tilde}) appears only after summing all the 
Kaluza-Klein tower of Pomeron pole terms.
}

One can also see that the saddle point expression (\ref{eq:GPD-saddle-approx})
automatically satisfies a requirement
that GPD should be consistent with DGLAP evolution,
because $\mu$-evolution is correctly taken into account in the $j$-plane expression (\ref{eq:GPD-in-j}).
This is a nontrivial requirement on GPD modeling in general.
One can consider, for example, a GPD profile given by
PDF (GPD at $t=0$) multiplied by some form factor
at a given renormalization scale \cite{Diehl:2005wq}:
\begin{equation}
\left[ \left(\frac{1}{x}\right)^{j^*} 
        \left(\frac{\Lambda}{\mu}\right)^{\gamma(j^*)} 
 \right] \times
 \frac{1}{\left(1 - B(x) t\right)^{p}},  
\end{equation}
where $B(x)=\alpha'(1-x)^3\ln(1/x)+\cdots$, and $\alpha'$ and $p$ are parameters.
The profile of GPD like this are not stable under DGLAP evolution.
On the other hand, 
 the GPD under the saddle point approximation (\ref{eq:GPD-saddle-approx}) is given
 by the PDF multiplied by a spin $j$ form factor 
 evaluated at the saddle point value $j=j^*$.
 The saddle point value $j^*$ depends on $x$ and the factorization/renormalization scale $\mu$.
This result obviously takes into account renormalization effects,
and hence is stable/reliable at any renormalization scale.

The remaining task is to determine the spin $j$ form factor 
at renormalized point $\mu=\Lambda$, $g^h_{i\nu_j}(\sqrt{-t}/\Lambda)$ 
as a holomorphic function of $j$.
This is along the line of the
collinear factorization approach (dual parametrization) to the modeling of GPD\cite{CF-DP}.
Derivation of $g^h_{i\nu_j}(\sqrt{-t}/\Lambda)$ from the first principle is an impossible task in perturbative QCD, because of the non-perturbative origin of the form factor,
and this is also hard in lattice simulation,
because there is practically no way of finding
 analytic continuation of integer spin matrix elements into complex $j$.
An alternative is to use predictions from the gauge/string duality,
and a crude way is to use the prediction of the hard wall model derived in
 \S3.1 as it is.
Indeed, as we saw, 
the hard wall model can explain decreasing
slope parameter $B$ of DVCS for large $q^2$,
 observed in HERA \cite{DVCS-slope-on-q}.
It is also possible to use more realistic gravity dual models for similar calculation, 
where at least we might 
want to require the model to have asymptotic free running for certain 
energy range (as in \cite{Aharony:1998xz}) still with large 't Hooft coupling.

If one wants to consider a gravity dual model that is truly dual to
the real world QCD (if there is any),
then it should run into a problem in its UV region
of the geometry because of large curvature.
This problem of gravitational description, however, may be alleviated
by borrowing the understanding of perturbative QCD.
Such strategy may not be totally nonsense.
We saw that
the singularities of the form factor in the $j$-plane are important in
 determining GPD,
and gravity dual with asymptotic free running suggests that 
the singularities are infinitely many poles\footnote{
These poles correspond to trajectories of Kaluza-Klein modes 
in radial direction $(z)$ of a single graviton trajectory in 10 dimensions
(or on $AdS_5$).
On top of this tower structure, there is yet another
tower structure of trajectories associated with the daughter trajectories
of stringy excitations on 10 dimensions.
}
 \cite{BrowerJHEP0712:0052007}.
The BFKL theory in perturbative QCD with a running coupling effect
also suggests infinitely many poles in the $j$-plane \cite{Forshaw:1997dc}.
Now, let us examine how sensitive the position of the poles 
 predicted from gravity
 dual are to 
the unreliable large curvature geometry in the UV region.
In gravitational descriptions,
each Pomeron pole has its wavefunction on the holographic coordinate,
and the Pomeron wavefunction becomes localized more and more
into the IR region of the holographic radius when ${\rm Re}\: j$ of the pole
 increases.
Therefore, the poles in large ${\rm Re}\: j$ are determined
 mainly by IR physics, and
position of poles predicted by gravity dual should be reliable,
while the poles in small ${\rm Re}\: j$ are 
quite sensitive to the unreliable geometry in the UV region.
As for such smaller ${\rm Re}\: j$ poles, however,
the position of the poles predicted by the BFKL theory 
(with asymptotically free running)
will be reliable.
Thus, by using both predictions from the gravity dual and the BFKL theory,
the poles in the $j$-plane may be properly determined.

In order to determine GPD completely,
not only the position of the poles but also complete profile of the spin $j$ 
form factor are required.
The spin $j$ form factor is given by integrating Pomeron wavefunction and impact factor
in gravity dual, and in fact, also in the BFKL theory;
the integration is carried out over the holographic radius $z$ in gravity dual,
whereas it is done over gluon transverse momentum $k_\perp$ in the BFKL theory.
The similar structure in the $k_\perp$ factorization formula and 
the gravity dual scattering amplitude (\ref{eq:Ii-def}) 
has been pointed out, and identification of 
gluon transverse momentum $k_\perp$ in the BFKL theory with
 holographic radius $z^{-1}$ in the gravity dual 
is suggested \cite{Brower:2002er, BrowerJHEP0712:0052007}.
Thus, one can retain the integration over the holographic radius 
in gravity dual in the IR (large $z$) region.
The integration in the UV region may be replaced by
that over $k_\perp$ coordinate in the BFKL theory;
this large $k_\perp$ region is where perturbative QCD is reliable.

 \section*{Acknowledgments}  

Part of this work was carried out during long term programs 
``Branes, Strings and Black Holes'' at YITP, 2009 (TW), 
``Strings at the LHC and in the Early Universe'' at KITP, 2010 (TW), 
``High Energy Strong Interactions 2010'' at YITP, 2010 (RN, TW) and 
also during a stay at Caltech theory group of TW. 
This work is supported by JSPS Research Fellowships for Young 
Scientists (RN), by WPI Initiative, MEXT, Japan (RN, TW) and 
National Science Foundation under Grant No. PHY05-51164 (TW).

\providecommand{\href}[2]{#2}
\begingroup
\raggedright

\endgroup

\end{document}